# A Global Appearance and Local Coding Distortion based Fusion Framework for CNN based Filtering in Video Coding

Jian Yue*, Yanbo Gao*, Shuai Li, Hui Yuan, *Senior Member, IEEE*, Frédéric Dufaux, *Fellow, IEEE*

*Abstract*—In-loop filtering is used in video coding to process the reconstructed frame in order to remove blocking artifacts. With the development of convolutional neural networks (CNNs), CNNs have been explored for in-loop filtering considering it can be treated as an image de-noising task. However, in addition to being a distorted image, the reconstructed frame is also obtained by a fixed line of block based encoding operations in video coding. It carries coding-unit based coding distortion of some similar characteristics. Therefore, in this paper, we address the filtering problem from two aspects, global appearance restoration for disrupted texture and local coding distortion restoration caused by fixed pipeline of coding. Accordingly, a three-stream global appearance and local coding distortion based fusion network is developed with a high-level global feature stream, a high-level local feature stream and a low-level local feature stream. Ablation study is conducted to validate the necessity of different features, demonstrating that the global features and local features can complement each other in filtering and achieve better performance when combined. To the best of our knowledge, we are the first one that clearly characterizes the video filtering process from the above global appearance and local coding distortion restoration aspects with experimental verification, providing a clear pathway to developing filter techniques. Experimental results demonstrate that the proposed method significantly outperforms the existing single-frame based methods and achieves 13.5%, 11.3%, 11.7% BD-Rate saving on average for AI, LDP and RA configurations, respectively, compared with the HEVC reference software.

## I. INTRODUCTION

With the boom of social media and short video centered applications, the amount of video data is growing dramatically, which requires more efficient video compression methods. Although High Efficiency Video Coding (HEVC) [1] has brought over 50% coding bit rate reduction compared to the H.264/AVC [2], it still uses the block-based video coding framework, incurring blocking and ringing artifacts in reconstructed video frames due to the coding inconsistency between coding units and quantization errors. To reduce such errors and blocking artifacts, filtering methods including the traditional de-blocking filter such as the deblocking filter [3] and sample adaptive offset (SAO) [4] and deep learning based ones have been developed [3], [5], [7]-[27], [34].

With the development of deep learning, many deep learning based in-loop filtering methods [5], [7]-[27], [34] have been proposed. Considering that in-loop filtering is a pixel-level restoration task to reduce the distortion at each pixel, many methods focus on exploring local features to preserve fine spatial details where the network modules process input and intermediate features at the input resolution. Advanced architectures have also been employed such as residual or dense networks [14], [20], [29], [31], [33], [36], [39], [40],[43] and attention mechanisms [24], [42], [59]. However, without the exploration of broad contextual information, desired fine texture information may not be able to be restored in the filtered image.

On the other hand, considering that the target of filtering is similar to image de-noising in terms of improving the quality of reconstructed frame, there are also methods employing de-noising neural networks as filtering in the video coding, which explores global semantic information such as the encoder-decoder [27], [28], [32], [44], [46] architecture with pooling and deconvolution. However, deconvolutions are not capable of generating spatially accurate outputs, and thus some video encoding-specific distortion such as blocking artifacts may not be well recovered.

Since the to-be-filtered reconstructed frame is obtained by a fixed pipeline of video coding procedures, including coding unit partition, prediction, transform coding and quantization, it is not only just a general distorted image but with specific video encoding distortion patterns. However, few efforts have been made to co-explore both features, i.e., to recover the global disrupted fine textures and deal with the specific local coding distortion. In order to address this problem, we propose a multilevel and mixed-scale feature fusion approach by

This work was partly supported by the National Key R&D Program of China (2018YFE0203900), the National Natural Science Foundation of China (No. 61901083 and 62001092) and SDU QILU Young Scholars program.

*indicates equal contributions.

J. Yue is with Shandong University, Jinan, China and University of Electronic Science and Technology of China, Chengdu, China. This work is done while J. Yue is visiting Shandong University.

Y. Gao, S. Li and H. Yuan are with Shandong University. Email: ybgao@sdu.edu.cn, shuaili@sdu.edu.cn, huiyuan@sdu.edu.cn.

F. Dufaux is with Université Paris-Saclay, CNRS, CentraleSupélec, Laboratoire des signaux et systèmes. Email: frederic.dufaux@l2s.centralesupelec.fr.

exploiting broad contextual information for restoring global appearance, and local coding distortion to recover fine spatial details.

The contributions of this paper can be summarized as follows.
1) The CNN based filtering in video coding is thoroughly investigated, and characterized from two aspects: restoring the disrupted texture in the reconstructed frame as a denoising process, and recovering the local video coding distortion resulted by a fixed pipeline of coding-unit based encoding procedures including prediction, transform and quantization. Experiments are designed to demonstrate that different pixels may be better filtered by different methods.
2) We present a global appearance and local coding distortion fusion framework for CNN based filtering in video coding, where three streams including low-level local feature extraction, high-level local feature extraction and global feature extraction are used.
3) An encoder-decoder architecture is used to extract the high-level global features and a mixed-scale residual block is developed to better capture the high-level local features. A channel and spatial attention based fusion (CSAF) is further developed to progressively fuse the global and local features for the final filtering.

Extensive experiments have been conducted with ablation studies on different modules. Experimental results demonstrate that the proposed method achieves state-of-the-art performance in All Intra coding configuration, and better performance than the existing single-frame based filtering methods in Low Delay and Random Access configurations. A preliminary version of this work has appeared in [55]. Significant improvements have been further made in this paper, where the mechanism of CNN based filtering is thoroughly investigated and a three-stream fusion framework is proposed. New modules have been developed to improve the performance of the streams and the final fusion including a mixed-scale residual block and channel and spatial attention based fusion (CSAF) method. More experiments are conducted including ablation studies to validate the effectiveness of the proposed method.

The rest of this paper is organized as follows. The related work on CNN based filtering in video coding is reviewed in Section II. Our proposed method is presented in Section III and the experimental results are shown in Section IV with ablation studies. At last, Section V draws the conclusion.

## II. RELATED WORK

Many deep learning based in-loop filtering methods have been proposed in the literature. In this section, we briefly describe the related work in three categories: local feature based, large scale feature based and coding information enhanced CNN filtering in video coding.

### A. Local feature based CNN filtering in video coding

Considering in-loop filtering is a pixel-wise generation/restoration task, many methods design CNNs operating on the original input image resolution in order to preserve the fine spatial details. In [5], a four layers convolution neural network named AR-CNN was proposed to deal with various artifacts in image compression. It directly processes the coded images in order to produce a high quality image. Considering directly generating an image is difficult than producing just the distortion, in [7] , a convolutional neural network named IFCNN was proposed, which uses a skip connection to add the input reconstructed frame to the output resulting in only producing the distortion with the network. In[10], a frame-based dynamic post-processing CNN was introduced using a 20-layer CNN model (VDSR-CNN) to extract feature from the reconstructed frame, and use side information of content complexity to assist the restoration of reconstructed frame. It is further enhanced by adding a classifier to adaptively choose whether different CUs use the VDSR-CNN based filtering [19]. In [14], [20], residual CNNs, dense CNNs and recursive CNNs were used, where deep networks are constructed for filtering. In[24], Squeeze-and-Excitation Filtering CNN (SEFCNN) was proposed for In-loop filtering, where a channel-wise attention network is used. In [25], an enhanced deep convolutional neural networks (EDCNN) was proposed with a weighted normalization and feature fusion block. A mixed loss with MSE and MAE is used for training. These methods all focus on local feature extraction without taking advantage of the global image information, thus mainly restoring the frame from the perspective of local coding distortion. In this way, some desired fine texture may not be kept in the filtered image.

### B. Large scale feature based CNN filtering in video coding

While in-loop filtering is a position-sensitive task where pixel-to-pixel correspondence from the input reconstructed frame to the output filtered frame is needed, the large scale semantic information can still assist the distortion recovering process. In [9], a variable-filter-size residual learning CNN (VRCNN) was further developed using filters with different sizes to capture features with different receptive fields. In addition to increasing the filter size, large scale features can also be extracted by increasing the receptive filed such as pooling. In [8],[26], CNNs with down-sampling and up-sampling (or deconvolution) blocks was used, where average pooling is adopted to enlarge receptive filed. In [11], a multi-scale convolutional neural network was proposed with an extra network branch of processing the down-sampled input features. In [35], a blind quality enhancement method (RBQE) was proposed to process different frames, e.g., easy or hard distorted frames, with a shallow (early exit) or deep subnet via a quality check procedure. These methods extract large-scale features to assist the filtering process. However, without proper exploration of the local coding distortion patterns, some spatial details could be lost, reducing the overall quality.

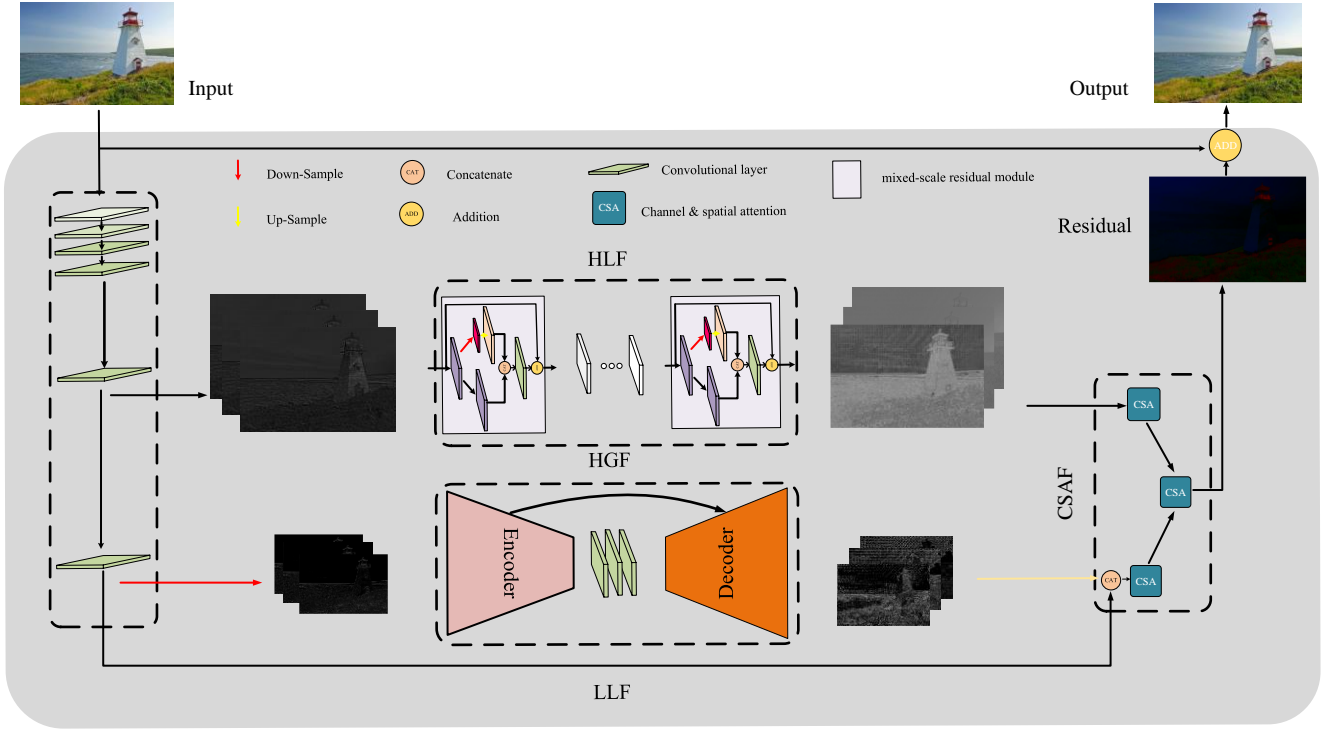

Fig.1. Proposed global appearance and local coding distortion based fusion framework. Three streams are used to extract low-level local feature (LLF), high-level global feature (HGF) and high-level local feature (HLF), and progressively fused by a channel and spatial attention based fusion (CSAF) module.

*C. Coding information enhanced CNN filtering in video coding*

In addition to the above methods focus on developing different neural networks to reduce the local coding distortion based on a single reconstructed frame, there are also methods aiming to explore more coding information in the filtering process. In [16], a partition-masked CNN was proposed using partition information as side information to unequally process the video contents based on their coding unit partition modes. In [15], the quantization parameter (QP) is used as extra input to make the single trained model applicable to reconstructed frames with all QPs. In addition to the coding information, there are also methods exploring multiple frames for filtering. In [13],[18],[21],[22], multiple temporally adjacent reconstructed frames or patches are used together for filtering where the temporal information is fused within the CNN framework. In [13], recurrent neural works such as LSTM are used for the temporal fusion of multiple frames. In [22], multiple high-quality reference frames, chosen by a reference frame selector (RFS), are first aligned by motion compensation net (MC-net) and then used to enhance the current frame. Similarly, in [34], the Kalman filter is employed for exploring multiple frames. With the coding information other than the input reconstructed frame explored for filtering, the performance can generally be improved. However, as investigated in the single-frame based methods [5,7-20,22,24,25], most of the information is contained in the reconstructed frame, and how to better extract useful features from the reconstructed frame is still the key problem for filtering, especially considering that in cases such as All Intra (AI) coding only the information from single frame is available for filtering. Moreover, it is worth investigating the single-frame based filtering as a basic and general module, which can be readily incorporated into temporal correlation enhanced filtering methods to further improve the performance. Therefore, this paper mainly investigates the CNN based video filtering using a single reconstructed frame and the exploration of other coding information can be extended upon the proposed method.

Overall, the existing methods either focuses on recovering spatial details but ignoring global semantical information with a single-scale pipeline, or uses large-scale features to explore broad contextual information but under-exploring the local coding distortion. Few efforts have been made to bring multilevel feature fusion to in-loop video filtering. However, the coded video frame is not only a distorted image but also distorted by a fixed pipeline of video encoding operations. The video in-loop filtering needs to consider both the global appearance by exploring the broad contextual information and the local coding distortion by taking advantage of the spatial details. Therefore, in this paper, we develop a multilevel feature fusion approach for CNN in-loop filtering by combing the global appearance and the local coding distortion information.

## III. PROPOSED APPROACH

In HEVC, the deblocking filter and SAO are used as the in-loop filters to reduce the coding artifacts. In this paper, we propose a CNN based in-loop filtering method to replace these conventional filters, which adaptively learns to reduce the compression artifacts in video encoding. The proposed method is described in the following.

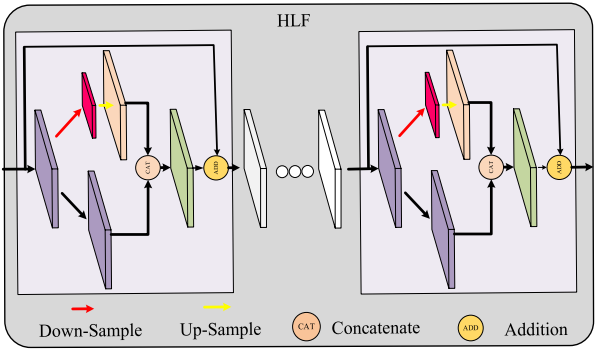

Fig.2 Mixed-scale residual block for high-level local feature extraction

*A. Proposed Global Appearance and Local Coding Distortion based Fusion Framework*

As briefly described in the Introduction Section, in the current video coding process, each frame is encoded by a fixed pipeline of operations including coding unit partition, intra/inter prediction, transform coding, quantization and entropy coding, then the reconstruction process (entropy decoding, inverse quantization, inverse transform and prediction compensation) to obtain the reconstructed frame $X \in R^{H \times W \times C}$ and then in-loop filter. $H$, $W$, and $C$ represent the height, width, number of channels of distorted image, respectively. The distortion in the reconstructed frames is introduced in the quantization process applied on transform coefficients (or directly the prediction residual if PCM mode is used), which may share similar patterns in different frames and videos. Such patterns mostly reside in each coding unit or the boundaries of neighboring coding units as the encoding process is performed on the basis of coding units. Therefore, local features are highly required and thus CNNs processing in the original resolution without pooling is used.

On the other hand, since the reconstructed frame is essentially a distorted image (not just coding units), the texture of the image also provides a strong prior for filtering. Therefore, the broad contextual information and high-level features are desired for restoring the unnatural and structurally destructive noisy frames. Accordingly, a CNN with capability of extracting high level features is also used in the proposed method.

To combing the advantages of contextualized high-level feature and local spatially-enriched features in restoration, we propose a mixed multilevel feature fusion approach for CNN in-loop filtering. The framework of the proposed approach is shown in Fig. 1, consisting of three CNN branches and an extra skip connection. The reconstructed video frame $X$ before filtering is used as the input. The extra skip connection adds the input reconstructed frame to the output to simplify the CNN to only produce the residual $R$ (restored distortion). The three branches extract the low-level local features $f_{ll}$, high-level local features $f_{hl}$ and high-level global feature $f_{hg}$, respectively, and are explained in the following subsections.

*B. Low-level local feature extraction (LLF)*

As noted above, local features learning the encoding-specific distortion is highly desired. In this paper, two branches are further used to obtain the low-level local features and high-level ones. The low-level local processing layers progressively produce features $f_{ll}$ with basic image semantics and the resolution of the features is held constant to support spatially-accurate mapping.

To be specific, a six-layer CNN without pooling was used for low level feature extraction as shown in the left branch of Fig. 1, and can be represented by

$$f_{out} = \sigma(\phi(W * f_{in} + b)), \quad (1)$$

where the $f_{in}$ and $f_{out}$ represent the input and output of each layer, respectively. Note that the input of the first layer in the network is the reconstructed frame $X$. The output of the network without the outer residual connection is distortion, which is of quite different characteristics from the input. Thus six-layers can be considered as a relatively shallow processing. Such low-level processing with the local features is able to capture the encoding distortion patterns within each coding units. The batch normalization [53] $\phi$ and ReLU activation $\sigma$ are used in each convolutional layer.

*C. Contextualized high-level global feature extraction (HGF)*

The video coding distortions, while appearing local patterns, also disrupts the image textures inevitably. The global texture of the reconstructed frame also can sever as a strong prior for restoration. For example, an object usually spreads across multiple adjacent coding tree units (CTUs) in video coding, especially in large resolution videos. The relationship between these blocks can be captured by global features and help the recovery of local losses within each CTU. In this paper, high-level features extracted with large receptive fields (using pooling) containing rich contextualized information is used as global information. To be specific, an encoder-decoder CNN similar to U-Net [37], [38] is used, which is the middle branch of the overall framework in Fig. 1. It consists of four progressive down-scale operations ($D\{\}$) with pooling to increase the receptive fields in order to yield the contextualized feature

$$d_{out} = D\{\sigma(\phi(W * d_{in} + b))\}, \quad (2)$$

where $d_{in}$ and $d_{out}$ represent the input and output of each down-scale operation $D$, respectively.

Four up-scale operations with up-sampling ($U\{\}$) is used to transfer the high-level global information back to high-resolution space and generate output $f_{hg}$. The intermediate high-level features extract the global information, and then the final high-resolution features produced together with the high-level global features facilitate the generation of the original resolution output

$$u_{out} = U\{\sigma(\phi(W * [u_{in}, d_{out}] + b))\}, \quad (3)$$

where $[u_{in}, d_{out}]$ and $u_{out}$ represent the input and output of each up-scale process.

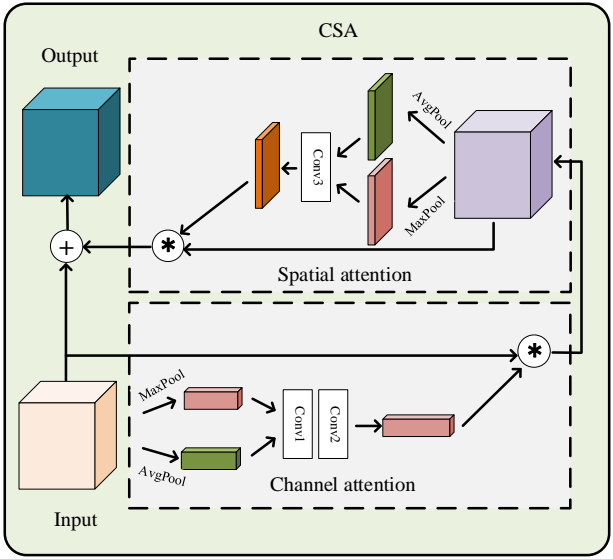

Fig.3. Channel and Spatial Attention (CSA)

At each down-scale process, max pooling of 2*2 is used to reduce the size of features and increase the receptive field. The number of channels is doubled after each max pooling. At each up-scale process, the opposite operation is applied with up-sampled resolution of features. The features obtained in the down-scale process are concatenated with the features of the same resolution in the up-scale process to guide the higher-level and higher-resolution feature extraction.

Such connections allow the subnet to preserve spatial features lost in down-sampling and also facilitates the back propagation of the gradient. In the down-scale and up-scale process, after each pooling or up-sampling, two convolutional layers are added to extract corresponding features. The filter size of the convolutional layer is 3x3, the channel number of the first convolutional layer is 64, and then increased or decreased according to the down-scale and up-scale operations. ReLU and batch normalization is used similarly as above.

### D. High-level local feature extraction (HLF)

As aforementioned in Subsection A and B, the reconstructed video frames are generated with a fixed procedures of coding operations, and the resulted distortions may share a similar property with rich spatial and local information. While the low-level local features may be able to represent the distortions in a coding units, it is difficult to extract features for a large neighborhood of coding units. Especially considering different coding units may be coded with different prediction modes and transform modes, the distortion patterns within each coding unit can be quite different from that of coding unit boundaries and a collection of coding units. Moreover, the down-sampling operation in the high-level global feature extraction branch may bring irreversible loss in detailed pixel information and this high-level local feature extraction branch can provide notion of spatial locations for the global features. In order to learn such high-level local features, a multiple mixed-scale residual module stacked CNN is developed as shown in Fig. 2. The mixed-scale residual module generates spatial features $f_{loc}$ on two different scales with an original CNN and an extra encoder-decoder block design.

Each mixed-scale residual module contains two branches, a trunk branch and a shallow encoder-decoder branch. Each trunk branch extracts features $x_t$ at its original resolution, i.e., local features,

$$x_t = \sigma(\phi(W_t * x_m + b_t)) \quad (4)$$

while the shallow encoder-decoder branch extracts features $x_e$ at a slightly larger scale by first down-sampling ($D\{\}$), processing and then up-sampling ($U\{\}$) the features

$$x_e = U\{\sigma(\phi(W_e * D\{x_m\} + b_e))\}, \quad (5)$$

This compensates the trunk branch to obtain locally larger-scale information such as information across neighboring coding units, and thus is concatenated with the features from the trunk branch. The mixed-scale residual module is shown in Fig. 2 and can be represented by

$$y_m = \sigma(\phi(W_m * [x_t, x_e] + b_m)), \quad (6)$$

At the end, a convolutional layer is further added to enhance the spatial features. A skip-connection is further added to facilitate the gradient backpropagation. Five mixed-scale residual modules are stacked to obtain the high-level local features $f_{hl}$.

### E. Channel and Spatial Attention based Fusion (CSAF)

To combine the low-level local features, high-level local features and high-level global features for both the global appearance-based and the local coding distortion-based filter, a progressive fusion framework is developed in this paper as shown in Fig. 3. Considering the low-level local features can bring spatial details to assist the high-level global features back-projected to the original resolution features, they are first concatenated together to obtain high-resolution features

$$f'_{hg} = \sigma(\phi(W * [f_{hg}, f_{hl}] + b)), \quad (7)$$

Then the high-level local features are further combined by a channel and spatial attention based fusion (CSAF) modified from the CBAM (Convolutional Block Attention Module) [41] using the spatial and channel attention to adaptively weight different features as shown in Fig. 3. The channel or spatial attentions are obtained by using max pooling and average pooling complementarily with pooling performed on spatial dimension or channel dimension, respectively. With the pooling operations, the complexity of a CSAF module is greatly reduced.

Different from [41], hyperbolic tangent activation (tanh) is used in our CSAF in order to learn both emphasizing or suppressing different features. Considering that with a residual connection in the attention block, the features can only be enhanced if using the sigmoid activation as in [41]. This may not be a problem for classification tasks as in [41] which uses relative features after softmax for final classification, but for video filtering providing precise pixel values, tanh could be more flexible. In our experiments, we also found that our CSAF with tanh consistently performs better than the original CBAM as shown in Section IV.C.3. Three CSAF modules are used in the proposed network as shown in Fig. 1, where CSAF is first applied for each branch and then fused together.

TABLE I
PROBABILITY OF GLOBAL NET AND LOCAL NET ACHIEVING BETTER
FILTERING RESULTS IN TERMS OF PIXEL PERCENTAGES

| Sequences | Probability |
|---|---|
| Class A | 51.83% |
| Class B | 51.51% |
| Class C | 51.66% |
| Class D | 50.75% |
| Class E | 52.32% |
| Average | 51.62% |

In the end, one last convolutional layer is used to produce the predicted coding distortion from the above enhanced features.

$$R = \sigma(\phi(W * CSAF(f_{hg}^{'}, f_{hl}) + b)) \quad (8)$$

Together with the input reconstructed frame added by a skip connection shown in Fig. 1, the final filtered output $Y$ can be obtained as $Y = X + R$.

Considering the first few layers in the three branches are of same resolution and function, they are shared together to reduce the number of parameters. The proposed global appearance and local coding distortion fusion network (GL-Fusion) can be trained end-to-end, and the mean squared error (MSE) was used as the objective function.

## IV. EXPERIMENTAL RESULTS

In this section, the experimental setups including the dataset establishment and implementation details are first introduced, and ablation studies on different modules are conducted to justify the development of the proposed GL-Fusion network. The proposed method is incorporated into the HEVC reference software (HM 16.9) and the performance is compared with state-of-the-art methods to validate the effectiveness of proposed method.

### A. Database Establishment

Considering that the proposed method investigates using only one single frame without other temporal frames, the AI (All-Intra) configuration in HEVC is used as the main comparison configuration. For AI, training based on images is the same to sequences without considering temporal correlation. Moreover, using images also increases the diversity of the training data since different frames in videos contain large portions of similar contents, making it redundant for the training. Therefore, for AI, the DIV2K [46] and Flickr2K [52] image datasets are used to establish our dataset for training. DIV2K and Flickr2K are high-quality (2K resolution) image datasets for image restoration or super resolution. DIV2K contains 900 high resolution images, and Fliker2K contains 2650 high resolution images. Firstly, the images are converted to YUV420 format and compressed under AI configuration by HM16.9. The native filters including SAO and De-blocking filtering are turned off in the compression. In addition to the main AI configuration, the LDP (Low-Delay-P) and RA (Random-Access) configuration in HEVC are also tested to demonstrate that the proposed method can also work well on other frame types. For LDP and RA, images cannot be used for training since no temporal correlation exists. Therefore, For LDP and RA, 108 raw YUV sequences collected from [49]-[51] are used for training. The sequences are compressed similarly as above only using the LDP and RA configuration, respectively. For LDP and RA, it is implemented as a postprocessing in the experiments, but it can be readily incorporated into the RDO optimization framework using the switchable filtering mechanism proposed in [24]. The sequences in the common test condition (CTC) by JCT-VC [1] are used for testing. The QP and other coding settings are the same to CTC including four QPs {22,27,32,37}.

TABLE II
COMPARISON OF DIFFERENT FEATURE EXTRACTION AND FUSION NETWORKS
(BD-RATE SAVING, %)

| Sequences | HGF | HLF | HGF+LLF | Fusion (LLF+HLF+HGF) |
|---|---|---|---|---|
| Traffic | -13.7 | -14.0 | -14.6 | -14.7 |
| PeopleOnstr. | -12.3 | -12.5 | -13.1 | -13.4 |
| Kimono | -12.1 | -12.3 | -12.9 | -13.0 |
| ParkScene | -10.4 | -10.7 | -11.2 | -11.2 |
| Cactus | -11.5 | -11.7 | -12.4 | -12.6 |
| BasketballD. | -12.0 | -12.4 | -13.1 | -13.7 |
| BQterrace | -7.6 | -7.7 | -8.2 | -8.4 |
| BasketballD. | -18.5 | -19.0 | -20.1 | -20.7 |
| BQmall | -12.0 | -12.7 | -13.1 | -13.4 |
| PartySence | -7.2 | -7.7 | -7.9 | -8.1 |
| RaceHorses | -8.0 | -8.1 | -8.6 | -8.9 |
| BasketballP. | -13.3 | -13.9 | -14.4 | -14.9 |
| BQSquare | -9.3 | -9.8 | -10.0 | -10.4 |
| BlowingB. | -9.9 | -10.3 | -10.6 | -10.9 |
| RaceHorses | -12.1 | -12.4 | -12.7 | -13.0 |
| FourPeople | -17.4 | -18.0 | -18.9 | -19.1 |
| Johnny | -17.2 | -18.3 | -18.8 | -19.5 |
| KristenAnd. | -16.3 | -17.2 | -17.9 | -18.1 |
| Class A | -13.0 | -13.6 | -13.8 | **-14.1** |
| Class B | -10.7 | -11.0 | -11.6 | **-11.8** |
| Class C | -11.4 | -11.9 | -12.4 | **-12.7** |
| Class D | -11.2 | -11.6 | -11.9 | **-12.3** |
| Class E | -17.0 | -17.8 | -18.5 | **-18.9** |
| Average | -12.30 | -12.70 | -13.20 | **-13.55** |

## B. Implementation Details

The proposed method is implemented based on PyTorch framework, and trained/tested on a Nvidia Tesla V100 card. Adam [48] is used as the optimizer and the initial learning rate is set to $2 \times 10^{-3}$. 4 models are trained for different QPs {22,27,32,37}. The patch size is set as 128*128, and the batch size is set to 32 in the training stage. At test stage, one entire frame is used as input for simplicity. The models are initialized using the normalization method in [47]. However, for the last layer producing the output distortion, the parameters are initialized differently. Since the output is directly used as the distortion not being normalized by softmax such as in classification problems, the output needs to be distributed in the range of the groundtruth distortion. Therefore, the parameters are initialized under the principle that the distribution of its outputs is the same to the groundtruth distortion. It is done by scaling the parameters with the standard variance difference between the actual output and the groundtruth distortion in a way similarly as in [47].

TABLE III
ILLUSTRATION OF THE PERFORMANCE WITH THE PROPOSED MIXED-SCALE RESIDUAL MODULE (BD-RATE SAVING, %)

| Sequences | Without the mixed-scale residual module | Proposed |
|---|---|---|
| Traffic | -14.0 | -14.7 |
| PeopleOnstreet | -12.6 | -13.4 |
| Kimono | -12.4 | -13.0 |
| ParkScene | -10.7 | -11.2 |
| Cactus | -11.9 | -12.6 |
| BasketballDrive | -12.8 | -13.7 |
| BQterrace | -7.5 | -8.4 |
| BasketballDrill | -19.7 | -20.7 |
| BQmall | -12.8 | -13.4 |
| PartySence | -7.7 | -8.1 |
| RaceHorses | -8.4 | -8.9 |
| BasketballPass | -14.2 | -14.9 |
| BQSquare | -9.8 | -10.4 |
| BlowingBubbles | -10.4 | -10.9 |
| RaceHorses | -12.6 | -13.0 |
| FourPeople | -18.2 | -19.1 |
| Johnny | -16.6 | -19.5 |
| KristenAndSara | -17.3 | -18.1 |
| Class A | -13.3 | **-14.1** |
| Class B | -11.0 | **-11.8** |
| Class C | -12.1 | **-12.7** |
| Class D | -11.8 | **-12.3** |
| Class E | -17.4 | **-18.9** |
| Average | -12.8 | **-13.55** |

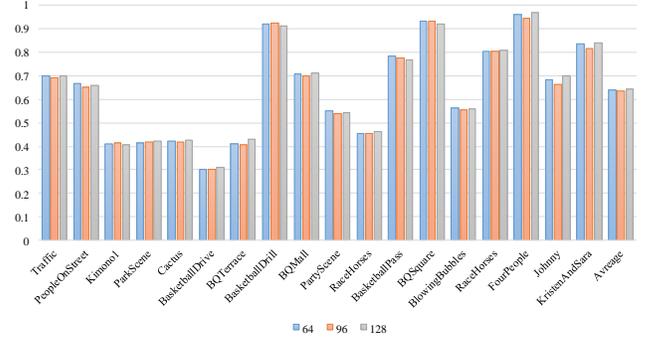

Fig.4. PSNR improvement comparison using different block sizes (64, 96, 128) for training.

## C. Ablation Studies and Analysis

In this subsection, extensive ablation studies are conducted to validate the proposed GL-Fusion modules including the necessity of contextualized high-level global feature extraction, high-level local feature extraction and the CSAF fusion, and different experimental settings such as the input size.

### 1) The necessity of the high-level global and local feature extraction

Ablation study on different networks are first conducted and the results are provided in Table II. It can be seen that the BD-rate saving [45] of the HGF, HLF and HGF+LLF is -12.3%, -12.7% and -13.2%, respectively, while it reaches -13.5% for the proposed GL-Fusion net. With the addition of the proposed features, the performance is consistently improved. On the other hand, to qualitatively illustrate the difference of different feature streams, we further designed an experiment to show the different effects brought by different features. First, a global feature stream consisting of LLF+HGF is designed for extracting broad contextual information desired for restoring the unnatural and structurally destructive noisy frames, corresponding to the proposed filtering aspect of global appearance based filtering. Note that LLF is added together with HGF in order to recover spatial details. Then the high-level local feature stream (HLF) is used for removing the specific coding distortion while preserving the spatially-enriched information, corresponding to the proposed filtering aspect of local coding distortion based filtering. To demonstrate that the two streams can complement each other and are both needed for filtering, experiments are conducted by setting each stream being one individual network.

The filtering results by each stream network are compared using their difference against the groundtruth frame $Diff = |y_g - y_t| - |y_l - y_t|$, where $y_g, y_l, y_t$ indicates the frames enhanced by global (with low-level local feature) net, high-level local net and the groundtruth frame, respectively. $Diff$ smaller than zero represent the global feature stream (HGF+LLF) performing better than the local feature stream on this pixel, and vice versa. Table I shows the percentage of global feature stream performing better, i.e., $Diff < 0$. It can

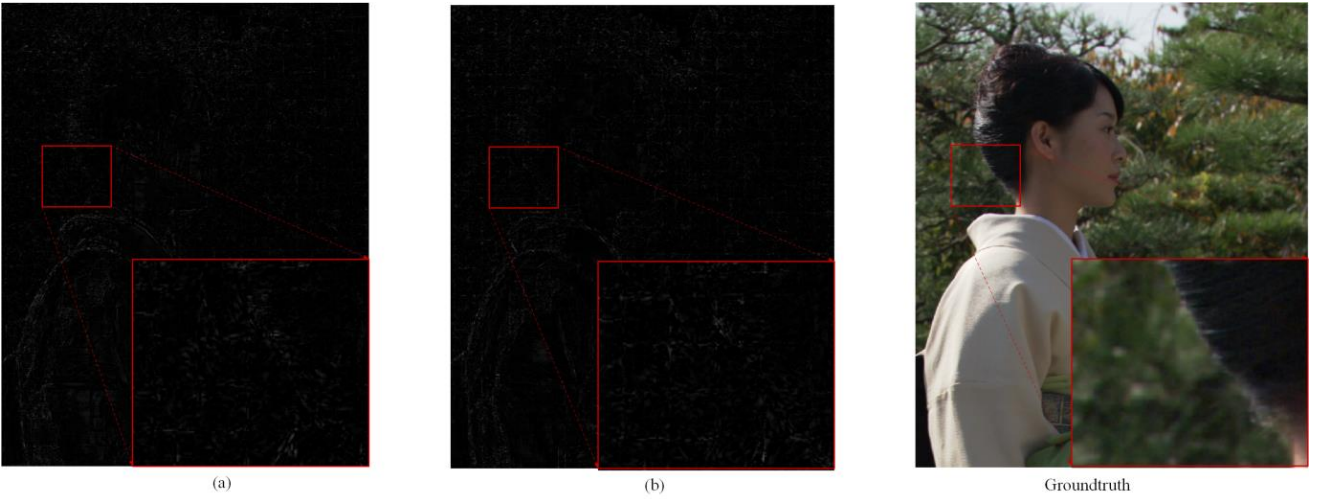

Fig. 5. Visualization of the absolute difference to the groundtruth of global feature stream net (a) and local feature stream net (b). The brighter the pixel (the values are transformed for better visualization), the better the enhancement. It can be seen that the local feature stream net can better handle the coding distortion around the coding unit boundary.

TABLE V
THE COMPARISON OF THE HYPERBOLIC TANGENT ACTIVATION AGAINST SIGMOID USED IN CSAF (BD-RATE SAVING, %)

| Sequences \ activations | Sigmoid | Hyperbolic tangent |
|---|---|---|
| Class A | -14.01 | **-14.05** |
| Class B | -11.06 | **-11.80** |
| Class C | -12.70 | **-12.74** |
| Class D | -12.21 | **-12.28** |
| Class E | -18.30 | **-18.87** |
| Average | -13.41 | **-13.55** |

TABLE IV
COMPARISON OF DIFFERENT FUSION METHODS (BD-RATE SAVING, %)

| Sequences | Addition | Concate. | SE. | CSAF |
|---|---|---|---|---|
| Traffic | -14.5 | -14.5 | -12.6 | -14.7 |
| PeopleOnstreet | -13.1 | -13.0 | -10.9 | -13.4 |
| Kimono | -13.0 | -12.8 | -11.1 | -13.0 |
| ParkScene | -11.2 | -11.0 | -9.8 | -11.2 |
| Cactus | -12.4 | -12.3 | -10.8 | -12.6 |
| BasketballDrive | -13.0 | -13.0 | -11.3 | -13.7 |
| BQterrace | -8.3 | -5.8 | -6.6 | -8.4 |
| BasketballDrill | -19.7 | -19.9 | -18.8 | -20.7 |
| BQmall | -13.2 | -13.0 | -12.3 | -13.4 |
| PartySence | -8.0 | -7.9 | -7.5 | -8.1 |
| RaceHorses | -8.6 | -8.6 | -7.3 | -8.9 |
| BasketballPass | -14.3 | -14.4 | -13.6 | -14.9 |
| BQSquare | -10.0 | -10.0 | -10.0 | -10.4 |
| BlowingBubbles | -10.6 | -10.6 | -9.8 | -10.9 |
| RaceHorses | -12.8 | -12.7 | -11.9 | -13.0 |
| FourPeople | -18.9 | -18.7 | -17.0 | -19.1 |
| Johnny | -19.1 | -16.7 | -17.0 | -19.5 |
| KristenAndSara | -17.9 | -17.6 | -16.3 | -18.1 |
| Class A | -13.8 | -13.7 | -11.7 | **-14.1** |
| Class B | -11.6 | -11.0 | -9.9 | **-11.8** |
| Class C | -12.4 | -12.4 | -11.5 | **-12.7** |
| Class D | -11.9 | -11.9 | -11.3 | **-12.3** |
| Class E | -18.6 | -17.6 | -16.8 | **-18.9** |
| Average | -13.25 | -12.90 | -12.00 | **-13.55** |

be seen that the percentage is around 50%, demonstrating that the two streams perform very similarly. Each stream can better handle half of the pixels than the other while worse for the other half (with different characteristics), agreeing with our statement that the two streams can complement each other. Therefore, the proposed GL-Fusion net with all the branches are necessary to achieve better performance on all pixels.

In Fig. 5, we also visualized the absolute difference to the groundtruth of the global feature stream net and local feature stream net. It can be seen that the local feature stream net can clearly handle the coding distortion near the coding unit boundary better, while the global feature stream focuses more on the edges of the contents/person and objects of the frame.

*2) Study on different block sizes in training*

The performance under different block sizes in training is also evaluated and the result is shown in Fig. 4. It can be seen that the average PSNR gain is 0.640, 0.630, 0.641 at block size 64, 96, 128, respectively. The performance is relatively similar and the block size 128 is used in our following experiments.

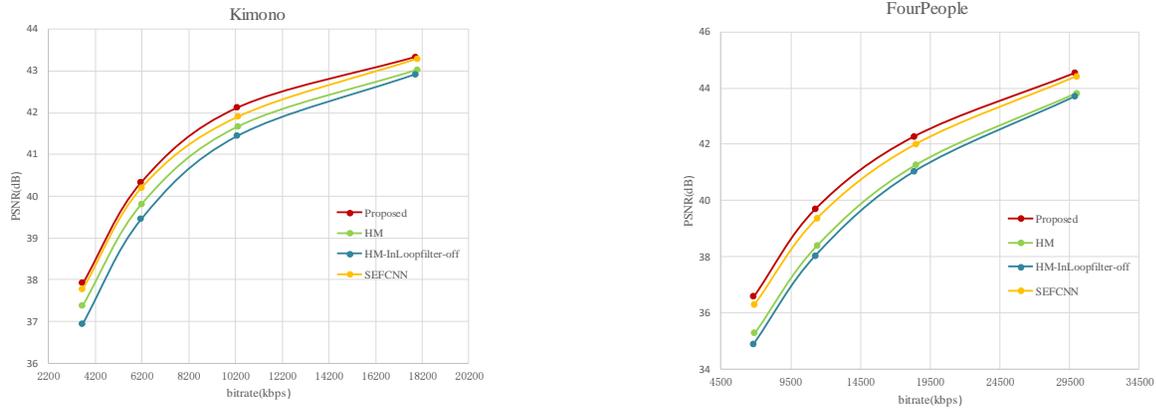

Fig.6. RD curves of the proposed method in comparison with the HM baseline and SEFCNN [24].

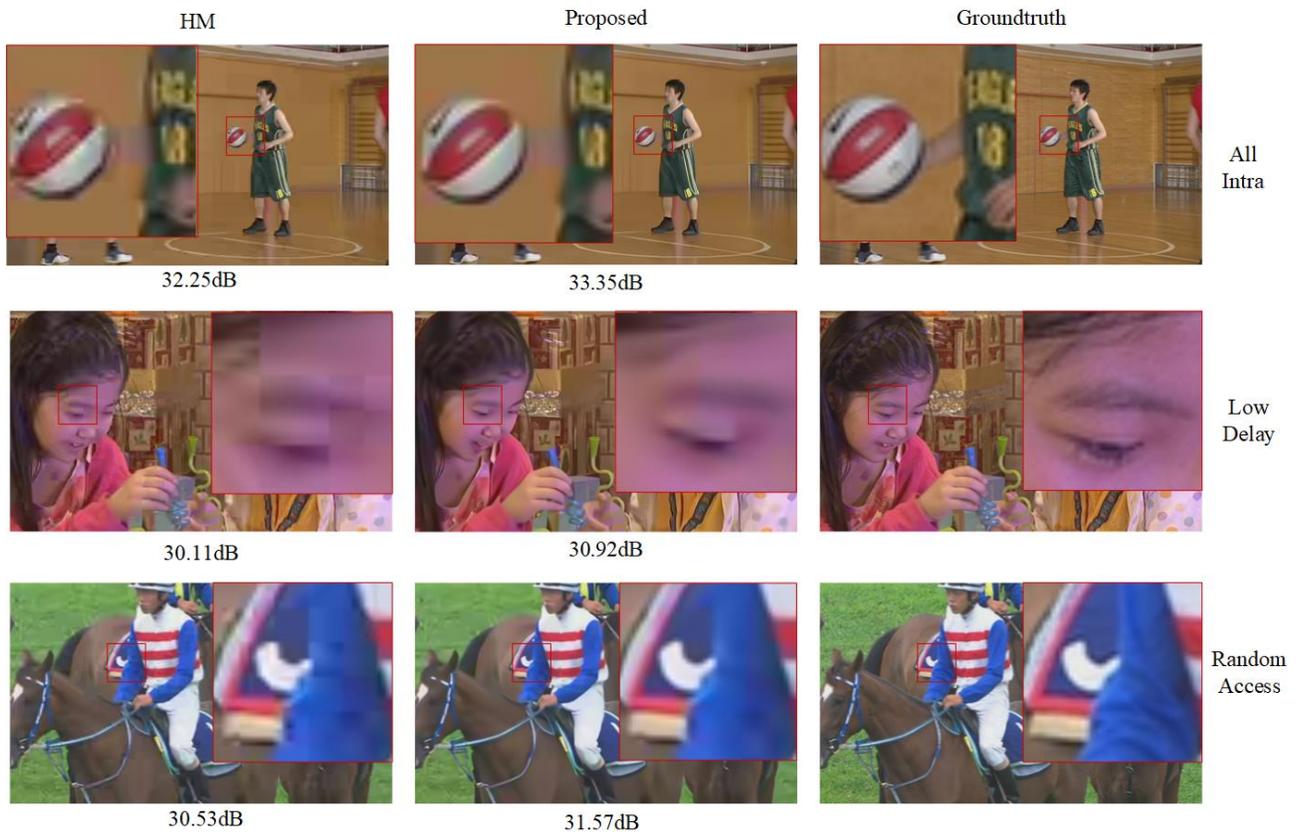

Fig. 7. Visualization of the filtered output by HM and the proposed method under different configurations

*3) Verify the effectiveness of the proposed mixed-scale local feature extraction*

To verify the effectiveness of the proposed mixed-scale local feature extraction, evaluation on the proposed mixed-scale residual module is conducted with comparison to the simple residual module (without the mixed-scale). The result is shown in table III. It can be seen that the proposed method with the mixed-scale residual module performs better on all the test sequences, validating the effectiveness of the mixed-scale local feature extraction.

*4) Exploration of different fusion schemes*

To evaluate the proposed channel and spatial attention based fusion (CSAF) for combining the global contextualized and local spatially-enriched features, several widely-used fusion schemes were compared, including addition, concatenation, and the squeeze and excitation (SE) [54]. The results are shown in table IV, and it can be seen that the proposed CSAF scheme achieves the best performance. Ablation study on using tanh or sigmoid for CSAF is also conducted and results are shown in Table V. It can be seen that the CSAF with tanh consistently performs better than sigmoid activation on all sequences with no extra cost.

TABLE VI
COMPARISON OF THE PROPOSED METHOD AGAINST THE STATE-OF-THE-ART METHODS (BD-RATE SAVING, %)

| Class | Sequences | Content Aware CNN [60] | | | MFRNET [61] | MIF [22] | | MRNN [23] | | VRCNN [9] | | | SEFCNN [24] | | | EDCNN [25] | | Proposed | | |
|---|---|---|---|---|---|---|---|---|---|---|---|---|---|---|---|---|---|---|---|---|
| | | AI | LD | RA | RA | LD | RA | AI | LD | AI | LD | RA | AI | LD | RA | LD | RA | AI | LD | RA |
| Class A (2560x1600) | Traffic | / | / | / | / | / | -6.80 | / | / | -5.61 | -1.81 | -3.54 | -11.21 | -6.84 | -7.48 | -5.02 | -5.97 | -14.73 | -14.30 | -12.30 |
| | PeopleOnstreet | / | / | / | / | / | -8.54 | / | / | -5.44 | -1.79 | -3.07 | -10.21 | -5.73 | -7.27 | -6.82 | -9.17 | -13.42 | -12.30 | -13.00 |
| Class B (1920x1080) | Kimono | / | / | / | / | / | -4.67 | -4.5 | -4.2 | -2.49 | -1.81 | -0.59 | -8.47 | -5.64 | -5.49 | -1.77 | -3.68 | -13.04 | -7.30 | -8.60 |
| | ParkScene | / | / | / | / | / | -4.32 | -5.1 | -4.8 | -4.41 | -0.96 | -1.82 | -8.57 | -3.56 | -4.76 | -2.77 | -3.85 | -11.20 | -7.40 | -8.70 |
| | Cactus | / | / | / | / | / | -8.29 | -7.9 | -8.4 | -4.54 | -3.79 | -5.10 | -8.17 | -7.13 | -7.98 | -3.38 | -5.29 | -12.60 | -11.10 | -12.90 |
| | BasketballDrive | / | / | / | / | / | -8.26 | -7.2 | -5.5 | -2.16 | -1.33 | -1.09 | -8.43 | -7.50 | -6.45 | -4.31 | -5.99 | -13.70 | -13.00 | -12.40 |
| | BQterrace | / | / | / | / | / | -10.0 | -8.3 | -9.2 | -2.62 | -6.02 | -4.66 | -5.48 | -11.45 | -9.20 | -13.08 | -5.08 | -8.40 | -8.90 | -12.70 |
| Class C (832x480) | BasketballDrill | / | / | / | / | / | -5.31 | -5.2 | -5.9 | -7.01 | -3.06 | -1.07 | -14.96 | -9.33 | -6.11 | -12.06 | -10.65 | -20.70 | -12.20 | -12.10 |
| | BQmall | / | / | / | / | / | -5.44 | -8.8 | -6.7 | -5.17 | -3.65 | -3.64 | -10.37 | -8.28 | -7.63 | -6.03 | -7.37 | -13.40 | -9.80 | -11.10 |
| | PartySence | / | / | / | / | / | -3.01 | -3.5 | -5.9 | -3.59 | -2.45 | -1.21 | -6.41 | -4.83 | -4.14 | -9.10 | -9.28 | -8.10 | -7.00 | -8.30 |
| | RaceHorses | / | / | / | / | / | -6.58 | -5.2 | -5.9 | -3.67 | -2.06 | -2.75 | -6.27 | -3.91 | -4.91 | / | / | -8.90 | -6.30 | -7.30 |
| Class D (416x240) | BasketballPass | / | / | / | / | / | -6.13 | -4.8 | -5.7 | -5.03 | -3.74 | -3.07 | -11.37 | -8.98 | -7.93 | -6.05 | -6.34 | -14.90 | -10.7 | -10.5 |
| | BQSquare | / | / | / | / | / | -4.30 | -9.2 | -9.1 | -3.92 | -4.50 | -0.81 | -8.13 | -9.38 | -6.61 | -13.08 | -10.99 | -10.4 | -11.60 | -13.70 |
| | BlowingBubbles | / | / | / | / | / | -3.10 | -5.0 | -7.5 | -4.88 | -2.62 | -2.19 | -8.53 | -4.96 | -4.28 | -8.10 | -7.54 | -10.90 | -8.90 | -9.20 |
| | RaceHorses | / | / | / | / | / | -6.55 | -4.4 | -4.0 | -7.08 | -2.06 | -5.30 | -10.75 | -6.09 | -7.61 | -3.27 | -3.52 | -13.00 | -9.40 | -10.0 |
| Class E (1280x720) | FourPeople | / | / | / | / | / | -9.40 | -10.6 | -15.1 | -7.01 | -5.77 | -7.02 | -14.76 | -13.79 | -13.88 | -8.85 | -10.53 | -19.10 | -17.80 | -16.50 |
| | Johnny | / | / | / | / | / | -8.77 | -8.6 | -14.4 | -5.98 | -4.80 | -3.49 | -13.67 | -14.63 | -12.42 | -9.83 | -11.22 | -19.50 | -19.10 | -16.60 |
| | KristenAndSara | / | / | / | / | / | -8.94 | -8.2 | -12.8 | -6.77 | -6.57 | -6.24 | -13.53 | -12.64 | -12.72 | -8.35 | -9.00 | -18.10 | -17.2 | -15.2 |
| | Class A | -4.7 | -3.5 | -6.6 | -7.87 | -7.24 | -7.67 | / | / | -5.52 | -1.80 | -3.03 | -10.71 | -6.29 | -7.38 | -5.92 | -7.57 | **-14.05** | **-13.25** | **-12.65** |
| | Class B | -3.5 | -4.5 | -6.5 | -7.57 | -5.84 | -7.11 | -6.6 | -9.7 | -3.24 | -2.78 | -2.22 | -7.82 | -7.06 | -6.78 | -5.06 | -4.78 | **-11.80** | **-9.50** | **-11.10** |
| | Class C | -3.4 | -4.4 | -4.5 | -9.51 | -4.16 | -5.08 | -5.7 | -6.1 | -4.86 | -2.80 | -2.17 | -9.50 | -6.59 | -5.70 | -9.06 | -9.56 | **-12.74** | **-8.80** | **-9.70** |
| | Class D | -3.2 | -3.5 | -3.3 | -13.05 | -3.49 | -5.01 | -5.9 | -6.6 | -5.23 | -3.70 | -2.84 | -9.69 | -7.35 | -6.61 | -7.63 | -7.10 | **-12.28** | **-10.20** | **-10.80** |
| | Class E | -5.8 | -7.7 | -9.0 | / | -8.15 | -9.04 | -9.5 | -14.1 | -6.59 | -5.77 | -5.58 | -13.99 | -13.69 | -13.00 | -6.34 | -10.25 | **-18.87** | **-18.00** | **-16.10** |
| | Average | -4.1 | -4.7 | -6.0 | -9.37 | -5.78 | -6.59 | -6.7 | -7.8 | -4.85 | -3.38 | -3.03 | -9.96 | -8.04 | -7.60 | -6.27 | -6.62 | **-13.55** | **-11.30** | **-11.70** |

### D. Comparison with State-of-the-art methods

The proposed method is compared against the state-of-the-art single-frame based CNN filtering methods, including MRRN [23], VRCNN [9], SEFCNN [24], EDCNN [25], Content Aware CNN [60], MFRNET [61], on All Intra, Low Delay or Random Access coding configurations. Moreover, MIF [22] as a basic multiple-frame based filtering is also compared to shed light on different categories of methods, but it is worth noting that we are mostly concerned with the comparison against the existing single-frame based filtering same to ours. The results are shown in table VI, and it can be seen that our proposed method achieves the best performance under all configurations with a relatively large margin (13.55% BD-Rate saving against 9.96% for the existing methods). Note that the methods are compared with the HEVC baseline with De-blocking and SAO on. RD curves on two example sequences are shown in Fig. 6, where the RD curve of the proposed method is compared with the HM and SEFCNN [24]. It can be clearly seen that the proposed method outperforms the existing methods. In addition, some qualitive results are also shown in Fig. 7. It can be seen that our method handles the blocking artifacts better than the deblocking and SAO filters in HEVC.

TABLE VII
TIME COMPLEXITY COMPARISON OF THE PROPOSED METHOD AGAINST HM16.9
BASELINE (TIME CONSUMPTION INCREASE, %)

| Sequences \ Configuration | AI | LD | RA |
|---|---|---|---|
| Class A | 14.99 | 5.79 | 4.09 |
| Class B | 16.62 | 11.55 | 2.56 |
| Class C | 14.67 | 10.68 | 7.60 |
| Class D | 14.08 | 3.56 | 8.18 |
| Class E | 17.90 | 17.29 | 8.56 |
| Average | **15.65** | **9.77** | **6.20** |

*E. Study of Computational Complexity*

The computational complexity of proposed method is compared with the HEVC standard reference software HM16.9 with deblocking filter and SAO on. The proposed method is incorporated into HM16.9, and NVIDIA Tesla V100 GPU is used for acceleration. As shown in Table VII, the proposed method takes 15.65%, 9.77%, 6.20% extra time consumption on AI, LDP, RA configuration, which is acceptable considering the large coding performance improvement.

V. CONCLUSION

This paper presents a global appearance and local coding distortion based fusion framework for CNN based filtering in video coding. It processes the filtering in video coding from two aspects: a denoising process where the reconstructed frame is a distorted image with disrupted texture, and a local restoration process where each coding unit is distorted by a fixed pipeline of encoding procedures including prediction, transform and quantization. Accordingly, a three-stream global appearance and local coding distortion based fusion network (GL-Fusion) is developed. One stream focusing on the contextualized high-level large-scale feature extraction for global texture restoration using a multilevel feature encoder-decoder. The second stream focusing on the high-level local feature extraction preserves spatially-enriched information for local coding distortion restoration resulted by the fixed pipeline of video coding operations. Mixed scale residual blocks are used to capture the local information. Finally, the last stream focusing on the low-level feature extraction to produce features with basic image semantics and the resolution of the features is held constant to the original resolution in order to support spatially-accurate mapping. The three streams are then progressively fused together by a channel and spatial attention based fusion to restore the original video. Experiments have shown that the proposed method outperforms the state-of-the-art result with 13.5% BD-Rate saving on average in All Intra configuration. The proposed GL-Fusion net can also be used and incorporated into the Versatile Video Coding (VVC) [56], which will be studied in future work.

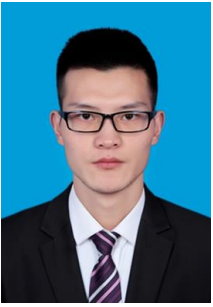

**Jian Yue** is currently pursuing his master's degree in School of Information and Communication Engineering, University of Electronic Science and Technology of China (UESTC), Chengdu, China. His research interests include video coding, deep learning and image processing.

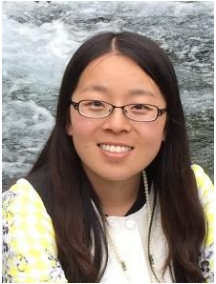

**Yanbo Gao** is currently with the School of Software, Shandong University (SDU), Jinan, China. She has been with the School of Information and Communication Engineering, University of Electronic Science and Technology of China (UESTC), Chengdu, China, as a Post-doctor since 2018. She received her Ph.D. degree from UESTC in 2018. Her research interests include video coding, 3D video processing and light field image coding. She was a co-recipient of a best paper award at the IEEE BMSB 2018.

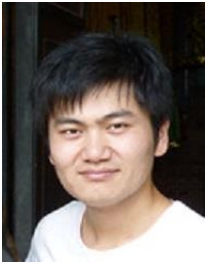

**Shuai Li** is currently with the School of Control Science and Engineering, Shandong University (SDU), China, as a Professor and QiLu Young Scholar. He was with the School of Information and Communication Engineering, University of Electronic Science and Technology of China, China, as an Associate Professor from 2018-2020. He received his Ph.D. degree from the University of Wollongong, Australia, in 2018. His research interests include image/video coding, 3D video processing and computer vision. He was a co-recipient of two best paper awards at the IEEE BMSB 2014 and IIH-MSP 2013, respectively.

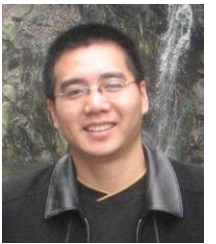

**Hui Yuan** (S'08–M'12–SM'17) received the B.E. and Ph.D. degrees in telecommunication engineering from Xidian University, Xi'an, China, in 2006 and 2011, respectively. In April 2011, he joined Shandong University, Jinan, China, as a Lecturer (April 2011–December 2014), an Associate Professor (January 2015–October 2016), and a Full Professor (September 2016). From January 2013–December 2014, and November 2017–February 2018, he was a Postdoctoral Fellow (Granted by the Hong Kong Scholar Project) and a Research Fellow, respectively, with the Department of Computer Science, City University of Hong Kong. His current research interests include video/image/immersive media processing, compression, adaptive streaming, and computer vision, etc. He served as an Area Chair and a workshop organizer of IEEE ICME2020.

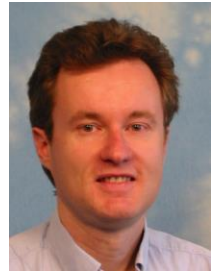

**Frédéric Dufaux** is a CNRS Research Director at Université Paris-Saclay, CNRS, CentraleSupélec, Laboratoire des Signaux et Systèmes (L2S, UMR 8506), where he is head of the Telecom and Networking hub. He received his M.Sc. in physics and Ph.D. in electrical engineering from EPFL in 1990 and 1994 respectively.

Frédéric is a Fellow of IEEE. He was Chair of the IEEE SPS Multimedia Signal Processing (MMSP) Technical Committee in 2018 and 2019. He is a member of the IEEE SPS Technical Directions Board. He was Vice General Chair of ICIP 2014, General Chair of MMSP 2018, and Technical Program co-Chair of ICIP 2019 and ICIP 2021. He is also a founding member and the Chair of the EURASIP Technical Area Committee on Visual Information Processing.

His research interests include image and video coding, 3D video, high dynamic range imaging, visual quality assessment, video surveillance, privacy protection, image and video analysis, multimedia content search and retrieval, and video transmission over wireless network. He is author or co-author of 3 books, more than 200 research publications and 20 patents issued or pending.